\documentclass[12pt]{article}
\usepackage{amsmath}
\usepackage{amssymb}
\usepackage{euscript}
\usepackage{epsfig}
\font\twelvess                  = cmss12

\font\twelvebm                  = cmmib10 at 12pt
\font\tenbm                     = cmmib10
\font\sevenbm                   = cmmib10 at 7pt

\mathchardef \be        = "0965
\mathchardef \bn        = "096E
\mathchardef \bp        = "0970
\mathchardef \bq        = "0971
\mathchardef \bt        = "0974
\mathchardef \bw        = "0977
\mathchardef \bx        = "0978
\mathchardef \by        = "0979
\mathchardef \bz        = "097A
\newcommand\Reals       {{\mathbb R}}
\newcommand\bzero       {{\hbox{\bf 0}}}
\newcommand\dC          {{\text{d}\esC}}
\newcommand\dS          {{\text{d}\esS}}
\newcommand\contv       {{x}}
\newcommand\thick       {{h}}
\newcommand\Varg        {{\alpha}}
\newcommand\Edist       {{\beta}}
\newcommand\Pt          {\mathop{\mbox{\twelvess pt}}\nolimits}
\newcommand\Tt          {\mathop{\mbox{\twelvess tt}}\nolimits}
\newcommand\Tp          {\mathop{\mbox{\twelvess tp}}\nolimits}
\newcommand\esA{{\EuScript A}}
\newcommand\esC{{\EuScript C}}

\newcommand\esP{{\EuScript P}}
\newcommand\esS{{\EuScript S}}

\font\big   = cmbx10 at 14pt

\begin{document}
\baselineskip = 2\baselineskip
\setcounter{page}{1}
\thispagestyle{empty}
         
%--------------
% Title / Author Info
%--------------

\begin{center}
{\big Self-interactions of strands and sheets}\\
\medskip
Jayanth R. Banavar${}^1$,
Oscar Gonzalez${}^2$, \\
John H. Maddocks${}^3$,
Amos Maritan${}^4$ \\
\end{center}

\bigskip

\begin{flushleft}
${}^1$Department of Physics, 104 Davey Laboratory, 
The Pennsylvania State University, University Park, 
Pennsylvania 16802 USA \\
\bigskip
${}^2$Department of Mathematics, The University of Texas, 
Austin, Texas 78712 USA \\
\bigskip
${}^3$ Institut Bernoulli, \'Ecole Polytechnique
F\'ed\'erale de Lausanne, CH-1015 Switzerland \\
\bigskip
${}^4$International School for Advanced Studies (SISSA),
Via Beirut 2--4, 34014 Trieste, Instituto Nazionale per la
di Fisica della Materia (INFM) and the Abdus Salam International
Center for Theoretical Physics, Trieste, Italy \\
\end{flushleft}

\vfill\eject

%--------------
% Abstract 
%--------------

\begin{abstract}
{\rm Physical strands or sheets that can be modelled as curves or 
surfaces embedded in three dimensions are ubiquitous in nature, 
and are of fundamental importance in mathematics, physics, biology 
and engineering.  
Often the physical interpretation dictates that 
self-avoidance should be enforced in the continuum model, 
i.e.\ finite energy configurations should not self-intersect.  
Current continuum models with self-avoidance frequently employ 
pairwise repulsive potentials, which are of necessity singular. 
Moreover the potentials do not have an intrinsic length scale 
appropriate for modelling the finite thickness of the physical 
systems.
Here we develop a framework for modelling self-avoiding 
strands and sheets which avoids singularities, and which provides 
a way to introduce a thickness length scale.  In our approach 
pairwise interaction potentials are replaced by many-body 
potentials involving three or more points, and the radii of certain 
associated circles or spheres.  Self-interaction energies based on 
these many-body potentials can be used to describe the statistical 
mechanics of self-interacting strands and sheets of finite 
thickness.}
\end{abstract}

\vfill\eject

%--------------
% Article Body
%--------------

\section{Introduction}
\label{SecIntro}

\begin{figure}
\centering
\epsfig{file=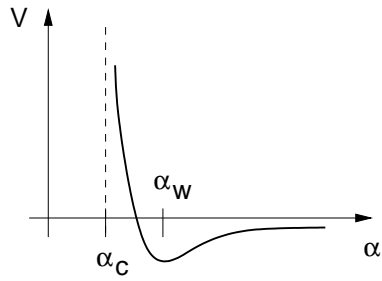}
\caption{ 
\baselineskip = 2\baselineskip
Example potential function with hard-core and potential well
parameters $\Varg_c$ and $\Varg_w$. }
\label{FigOne}
\end{figure}

The physical world consists of interacting matter. Many systems
arising in science can be modelled effectively with a finite number 
of distinct constituent particles or point masses, $\bq_i\in\Reals^3$
($i=1,\dots,N$).  A rich variety of natural phenomena -- such as the
existence of distinct phases (i.e.\ solid, liquid, and gas), and
transitions between them -- may be understood merely on the basis 
of pairwise interactions in such systems, with the phenomenological
pairwise interaction potential being induced by more fundamental
interactions at the atomic level. Many-body interactions between
triplets or quadruplets of particles are usually only treated as 
a higher-order correction. When the particles are all identical,
it suffices to introduce a single potential energy function $V(\Varg)$ 
that is defined for all scalar arguments $\Varg>0$. Then the potential 
between distinct particles with labels $i$ and $j$ is given by 
$V(|\bq_i - \bq_j|)$, where $|\cdot|$ denotes the usual Euclidean 
norm. We will be particularly concerned with systems where the 
potential has a qualitative form
akin to that depicted in Figure \ref{FigOne}, i.e.\ strongly repulsive 
or infinite at short distances, and with a well at some finite distance
$\Varg_w$. Provided that the potential is sufficiently repulsive in the
sense that $V(\Varg)\to\infty$ as $\Varg\to 0$, all finite energy
configurations have distinct locations for all particles, i.e.\
$\bq_i\neq\bq_j$ for all distinct $i,j=1,\dots,N$. Consequently, in
any finite energy configuration, the minimum over all pairwise
distances $|\bq_i-\bq_j|$ yields a distance of closest approach for
the configuration that is positive.  And it can be anticipated that
the location $\Varg_w$ of the potential well provides a characteristic
length scale for this distance of closest approach.

We shall consider here the case in which the interacting particles are
not unconstrained, but are instead restricted to lie on, or close to,
a $D$-dimensional manifold embedded in $\Reals^3$, such as a curve
($D=1$) or a surface ($D=2$).  Such systems are widely studied in many
branches of science; examples include polymers
\cite{YAMAKAWA}--\cite{CLOISEAUX}, and random surfaces or membranes
\cite{NELSON}, \cite{WIESE}.  The potential that gives rise to the 
internal stresses that maintain the system close to a continuous
manifold is not our concern here -- there are many such tethering
potentials for curves and surfaces \cite{GENNES}--\cite{WIESE} that are 
entirely satisfactory.  Rather our focus is on the self-interactions that 
can arise when the curve or surface is sufficiently deformed in $\Reals^3$
so that very different parts of the manifold come together to form, or
are close to forming, a self-intersection. Such self-interactions are
not captured by the standard local tethering potentials, and an
additional non-local self-interaction potential must be
introduced. Our key result is that in order to have a singularity-free
description in the continuum limit, the self-interaction potential
must depend on $D+2$ or more points.  That is, for curves a
non-singular self-interaction potential must depend on three or more
points (and such non-singular interaction potentials exist), while for
surfaces non-singular self-interaction potentials must depend on at
least a four-point interaction.  Notice that within this framework the
standard case of a discrete number of unconstrained particles
corresponds to $D=0$, so that pairwise interactions suffice.

The usual descriptions of a self-avoiding curve or surface employ 
pairwise potentials that are singular and which lack a length scale that 
can be identified with the physical thickness of the system.   
For example, a single uniform strand of cooked spaghetti 
of length $L$ and thickness $\thick$ would typically be modelled 
by a curve $\esC$, which may be interpreted as the centerline of the 
spaghetto, together with an effective potential energy 
\cite{DOI}, \cite{CLOISEAUX}
\begin{equation}
E_2[\esC] =   \int_0^L U(\bq(s),\bq'(s),\ldots)\;\dC_s
 +\nu \int_0^L\int_0^L  \delta(\bq(s)-\bq(\sigma)) \;
\dC_s\;\dC_\sigma
\label{Ham_2}
\end{equation}
where $U$ is a specified function, $\bq(s)$ is a parameterization of
the curve, $\dC$ is an element of arclength, $\delta$ is the Dirac
delta function, and $\nu$ is a model parameter.  The first term arises
from tethering effects at the microscopic level, and yields effective
internal stresses in the strand.  The second term, with its singular
pairwise potential, models self-avoidance of the strand. But this
model does not capture effects of the physical thickness parameter
$\thick$, which indeed does not even appear.  The analysis and numerical
treatment of such singular models is plagued by divergences that can
only be handled using sophisticated mathematical techniques
\cite{WILSON}, \cite{ZINN}.

\begin{figure}
\centering
\epsfig{file=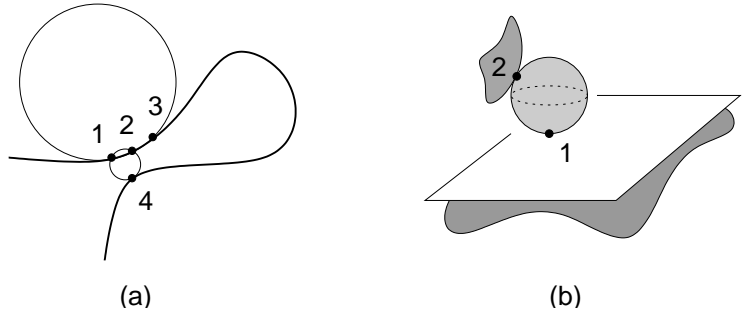}
\caption{
\baselineskip = 2\baselineskip
Interpretation of multi-point distances for a curve and a
surface.  ({\bf a}), Three-point distance for curve. Given any three
distinct points, $r$ is the radius of the unique circle that contains
the points.  When the points are from the same neighborhood on a
curve, such as points 1, 2 and 3, $r$ is close to the local radius of
curvature.  When points, such as 1, 2 and 4, are taken from two
different neighborhoods of the curve that are close to intersection,
$r$ approximates (half of) the distance of closest approach of the
curve to itself.  ({\bf b}), Tangent-point distance for a surface.
Given two distinct points 1 and 2 on a surface, $\Tp$ is the radius of 
the unique sphere that contains both points and is tangent to the surface 
at point 1.  When the points are neighbors on the surface, $\Tp$ approximates
the absolute value of the local normal radius of curvature in the
direction defined by the two points (not illustrated).  When points are 
taken from different neighborhoods that are close to intersection,
$\Tp$ approximates (half of) the distance of closest approach.  }
\label{FigTwo}
\end{figure}

In this article we develop models of self-avoiding curves and surfaces
that are nonsingular, and which include an explicit thickness length
scale.  In particular, we replace the singular delta-function
potential above by a smooth potential $V(\Varg)$ of the qualitative 
form shown in Figure \ref{FigOne} that is dependent upon a single
scalar variable $\Varg$.  The only difference between our models for
discrete sets of points, for curves, and for surfaces, will be our
choice for the independent variable $\Varg$.  Moreover, our continuum
models retain the following two desirable features of the
unconstrained discrete case described above: i) all finite-energy
configurations of the manifold are non-self-intersecting, with a 
positive distance of closest approach, and ii) the location $\Varg_w$ 
of the well in the potential $V(\Varg)$ provides a characteristic
length scale for the distance of closest approach of the manifold.
The difficulty in constructing such models is that the standard choice
of taking the argument $\Varg$ of the potential to be the Euclidean
pairwise distance cannot satisfy our two desiderata when the
underlying system is continuous. The reasons are discussed more
precisely in Section \ref{SecCurv} below, but the basic idea is simple. 
For the curve depicted in Figure \ref{FigTwo}(a), one wishes to penalize 
true points of closest approach between distinct parts of the curve, as, 
for example, between points $1$ and $4$, {\em without} penalizing adjacent 
points from the same part of the curve, as for example points $1$ and $2$.
The pairwise Euclidean distance simply cannot distinguish between
these two cases; in other words, it cannot distinguish between
proximity of points that is forced by continuity of the manifold in
{\em any} configuration, and the real phenomenon of interest, namely
proximity of points due to large scale geometrical deformation in {\em
some} configurations.

Our principal result is that the above two desiderata can be achieved
simply by taking the argument $\Varg$ to be a quantity other than the 
Euclidean distance between two points. Specifically, as discussed in 
Section \ref{SecCurv}, for the case of curves $\Varg$ can be taken to 
be the radius of the circle defined by three points.  When the argument 
of $V(\Varg)$ depends upon more than simply two points, for 
example triplets or quadruplets of points, we shall refer to $V(\Varg)$ 
as a many-body or multi-point potential. The use of many-body potentials 
is an essential ingredient in the models of continuous systems that we
propose, and they should not be viewed as a higher-order correction to
two-body or pairwise potentials, as is the case in discrete
models. Indeed our proposal for continuous models is to {\em replace}
pairwise self-interaction potentials, which must be singular, with
many-body self-interaction potentials, which need not be singular.
For example, for the spaghetto problem, we suggest an effective
energy of the form
\begin{equation}
E_3[\esC] =   \int_0^L U(\bq(s),\bq'(s),\ldots)\;\dC_s
 + \int_0^L  \int_0^L \int_0^L V(r(s,\sigma,\tau)) \;
\dC_s\;\dC_\sigma\;\dC_\tau
\label{Ham_3}
\end{equation}
where the last term in the more standard energy (\ref{Ham_2}) has been
replaced by a three-body potential $V(r(s,\sigma,\tau))$ with $V$ of
the form described in Figure \ref{FigOne}.  Here $r(s,\sigma,\tau)$ is 
the radius of the circle defined by the three points $s$, $\sigma$ and 
$\tau$ as discussed in Section \ref{SecCurv}.  The repulsive potential 
in (\ref{Ham_3}) can be finite because in the limit of three points
coalescing along the curve, for example points $1$, $2$ and $3$ in
Figure \ref{FigTwo}(a), the radius $r$ tends to the local radius of 
curvature, which is well-defined and positive for twice continuously
differentiable curves.  (A more detailed discussion of smoothness 
assumptions is given in Section 2.) However, whenever different parts 
of a curve come together to form a self-intersection, there are triplets 
of points, such as $1$, $2$, and $4$ in Figure \ref{FigTwo}(a), for 
which the corresponding circle radius $r$ approaches zero.  
The self-avoidance of the uniform spaghetto of thickness $h$ is 
modelled by a hard-core potential $V$ which is infinitely large when 
its argument is less than $h$ and zero otherwise.  Likewise, as 
described in Section \ref{SecSurf}, a certain generalized four-body 
potential can be used to model self-avoiding surfaces of finite 
thickness.

It should be stressed that the difficulties we address simply do not
arise in a many-body system with a discrete index for the particles.
On the other hand, analytic treatments of interacting systems with a
very large number of particles are often facilitated by making a
continuum approximation, in which the discretely indexed particles
$\bq_i$ are replaced by a field $\bq(\contv)$ that is dependent upon a
continuously varying independent variable $\contv$. 
The appropriate phenomenological interaction potential for the 
continuum description is to be derived from the microscopic ones.  
In the case of continuous phase transitions, this is a powerful
procedure because the critical behavior is unaffected by the precise
microscopic interactions \cite{KADANOFF}, \cite{STANLEY}, and the 
critical exponents can be derived using field-theoretic techniques
\cite{WILSON}, \cite{ZINN}. But it is the passage to the continuum limit which
implies that any two-body self-interaction potential must be singular. In
order to avoid such singularities we suggest that in the continuum limit
the effective potentials modelling non-local self-interactions should
be many-body ones.

\section{Self-interactions of curves}
\label{SecCurv}

We consider various distances, other than the usual Euclidean pairwise
one, that can be associated with points on a curve.  Here and
throughout a curve $\esC$ will mean a function $\bq(s)\in\Reals^3$ of
a variable (arc-length) $s\in [0,L]$. We shall consider only
sufficiently smooth curves, specifically those that are twice
continuously differentiable.  This stands in contrast to some models
in field theory where polymers are sometimes represented by curves
that are continuous but not smooth, for example piecewise linear.  In
point of fact there is an emerging body of literature
\cite{CKS}--\cite{SCHURICHT} which suggests that our ideas could
usefully be applied to a slightly larger class of curves, namely those
with only a Lipschitz continuous first derivative, but we do not
pursue such questions here. A curve will be called {\em simple } if it
has no self-intersections; that is, if $\bq(s)=\bq(\sigma)$ only when
$s=\sigma$.

In \cite{KATRITCHa} it was shown that certain {\em ideal} shapes of
knots are related to various physical properties of knotted
DNA. Intuitively these ideal configurations can be described as having
the property that for a given knot-type and prescribed length they are
as far as possible from self-intersection. The idea of a three-point
distance based on the radius of the associated circle was introduced
in \cite{GONZALEZ} as one way to make the notion of distance from
self-intersection mathematically precise. In \cite{MARITAN} the same
three-point circular distance was used as an ingredient in the
numerical study of the optimal shapes of compact strings. The
properties and relations between all possible circular and spherical
distance functions defined on curves are discussed at length in
\cite{smutny}.

In the present article we argue that these generalized circular and
spherical distances also provide natural means for defining
singularity free self-interaction energies of curves through a potential
function $V(\Varg)$ with the qualitative form of Figure \ref{FigOne}
that takes a multi-point distance as argument. For our purposes, a
self-interaction energy will mean a functional $E[\esC]$ that is
finite for any simple curve $\esC$, and which tends to infinity as
$\esC$ tends to a non-simple curve. We remark that within the specific
context of knot theory the construction of simple geometric
self-interaction energies for curves has already received much
attention; the case of a pure inverse power of a circular three-point
distance was proposed in \cite{GONZALEZ}, and surveys of alternative
approaches can be found in several Chapters of \cite{STASIAK}.

\subsection{Two-point distance for curves}

Given an arbitrary simple curve $\esC$, the most intuitive approach to
the construction of a scalar argument $\Varg$ for a self-interaction
energy $V(\Varg)$ is to take the usual pairwise or two-point distance
function
$$
\Edist(s,\sigma)=|\bq(s)-\bq(\sigma)|.
$$
In particular, a candidate energy 
functional $E[\esC]$ would then be the double integral 
$$
E[\esC] = \iint V(\Edist(s,\sigma)) \;\dC_s\;\dC_\sigma
$$
with for example $V(\Varg)=\Varg^{-m}$.  The basic idea is that, 
for $m\ge2$, the integral tends to infinity
as $\bq(s)$ tends to $\bq(\sigma)$ with $s\ne \sigma$, thus meeting
the infinite-energy condition associated with self-intersections.
However, such an integral is {\em always} divergent due to
nearest-neighbor effects since $\Edist(s,\sigma)=0$ when $s=\sigma$, so
that the energy $E[\esC]$ is infinite for {\em any} curve $\esC$.

To cure the above divergence problem one may consider regularizing 
the integrand by subtracting something equally
divergent as $s\to\sigma$, or mollifying the integrand using a
multiplicative factor that tends to zero at an appropriate rate as
$s\to\sigma$. (See for example \cite{STASIAK} and references therein.)
In essence, these procedures introduce a length scale to compensate
for the fact that there is no inherent small-distance cutoff for the
pairwise distance between nearest-neighbors along a curve.
Renormalization group techniques may then be used to extract critical
behavior that is independent of this artificial cutoff length scale
\cite{CLOISEAUX}, \cite{ZINN}.

\subsection{Three-point distance for curves}

An alternative approach to defining the argument $\Varg$ of the
self-interaction energy is based on triples of points.  To begin,
consider any three distinct points $\bx$, $\by$ and $\bz$ on a simple
curve $\esC$.  When these points are not collinear they define a
triangle with sides of lengths $|\bx-\by|$, $|\bx-\bz|$ and
$|\by-\bz|$, perimeter $\esP(\bx,\by,\bz)$ and area
$\esA(\bx,\by,\bz)$.  Each of these quantities vanishes in any limit
in which all the points coalesce into one, so they do not individually
yield an appropriate length scale for self-interaction. On the other
hand, certain combinations provide quantities that remain positive in
coalescent limits.  One natural combination is
$$
r(\bx,\by,\bz) = \frac{|\bx-\by||\bx-\bz||\by-\bz|}
                                    {4\esA(\bx,\by,\bz)}
$$
which can be identified as the radius of the circumcircle, i.e.\ the
unique circle passing through $\bx$, $\by$ and $\bz$.  By convention, 
we take this radius to be infinite when the points are collinear.

Various properties of the three-point circumradius function
$r(\bx,\by,\bz)$ were studied in \cite{GONZALEZ}.  For our 
purposes we merely note that the domain of the function $r(\bx,\by,\bz)$ 
can be extended by continuous limits to all triples of points on $\esC$,
distinct or not.  For example, if $\bx=\bq(s)$, $\by=\bq(\sigma)$ and
$\bz=\bq(\tau)$ are three distinct points on $\esC$, then it is
straightforward to show that
\begin{equation}
\lim_{\sigma,\tau\to s}r(\bx,\by,\bz) = \rho(\bx)
\label{ThrPntLim}
\end{equation}
where $\rho(\bx)$ is the standard local radius of curvature of $\esC$
at $\bx$.  (Because we consider only curves that are twice
continuously differentiable this coalescent limit exists at each
point; see \cite{GSMVdM} for further details when the underlying curve
is not smooth.)  From its geometrical interpretation we may also
deduce that $r(\bx,\by,\bz)$ and its limits are invariant under
translations and rotations of a curve.  Moreover, whenever different
parts of a curve come close to forming a self-intersection, there are
points $\bx$ and $\by$ for which the limits $r(\bx,\by,\by)$ and
$r(\by,\bx,\bx)$ are both equal to half of the distance of closest
approach.

A general class of self-interaction energies involving the 
circumradius function can now be defined.  In particular,
one may consider the energy 
$$
E[\esC] = \iiint V(r(s,\sigma,\tau)) 
              \;\dC_s\;\dC_\sigma\;\dC_\tau
$$
where
$$
r(s,\sigma,\tau)=r(\bq(s),\bq(\sigma),\bq(\tau))
$$
and $V$ is of the form introduced in Figure \ref{FigOne}, or more 
simply $V(\Varg)=\Varg^{-m}$ with $m$ an appropriately large exponent.  
In contrast to the pairwise distance function $\Edist(s,\sigma)$, the
circumradius function $r(s,\sigma,\tau)$ does not suffer
nearest-neighbor effects due to curve continuity.  In particular,
$E[\esC]$ is well-defined for any simple curve $\esC$. 

The condition $V(\Varg)\to\infty$ as $\Varg\to 0$ simultaneously 
provides control over a curve at both a local and global level.  
For example, at a point $s$ on the curve, the limit $r(s,s,s)$ 
defined in (\ref{ThrPntLim}) is just the local radius of curvature 
at that location and the potential $V(r(s,s,s))$ 
is finite as long as this radius is non-zero (or equivalently, 
as long as the local curvature is finite). In this case, the 
three-body potential $V$ plays the role of a local curvature 
energy that encourages curve smoothness.   On the other hand, 
whenever different neighborhoods of a curve come together to form 
a self-intersection, there are points of closest-approach 
$(s,\sigma)$ for which the limits $r(s,\sigma,\sigma)$ 
and $r(\sigma,s,s)$ tend to zero, leading to infinite values for 
$V(r(s,\sigma,\sigma))$ and $V(r(\sigma,s,s))$. Thus, the 
three-body potential is also a global self-interaction energy 
that acts to discourage self-intersections.  

Characteristic length scales for curve self-interactions can be
identified depending on the functional form of $V$. For example,
potentials with the general form introduced in Figure \ref{FigOne} 
provide a natural scale for modelling the steric self-interactions 
of material filaments with non-zero thickness.  A single generic 
potential (repulsive at short $r$, attractive for intermediate $r$) 
based on a three-body argument $r(s,\sigma,\tau)$ suffices to obtain 
both a swollen phase at high temperatures and a dense phase at low
temperatures, along with a phase transition between them in analogy
with the fluid-solid transition for unconstrained particles with a
pairwise potential.  In the more standard continuum approach
\cite{GENNES}--\cite{CLOISEAUX} a similar transition is obtained by
introducing both attractive singular two-body (in order to encourage
collapse) and repulsive singular three-body (in order to account for
self-avoidance) $\delta$-function types of potentials.

\begin{figure}
\centering
\epsfig{file=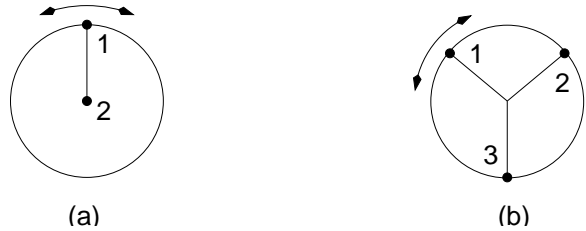}
\caption{ 
\baselineskip = 2\baselineskip
Graphical interpretation of forces derived from potentials
depending on two- and three-point distance functions.  ({\bf a}), two
particles interacting via a potential depending on two-point or
pairwise distance.  Holding particle 2 fixed and moving particle 1 on
the circumference of the circle leaves the
energy unchanged. This implies that the force on each particle is along the
joining line, and that the resultant of the forces is zero due to the
translational invariance of the energy.  ({\bf b}), three particles
interacting via a potential depending on three-point distance.  Here
any combination of the particles may be moved along the circumference
of the circle without changing their interaction energy.  This implies
that for any distinct triplet the force on each particle is along a radial 
line from the center of the circle to the particle, and that the resultant 
of the forces is again zero due to the translational invariance of the energy.}
\label{FigThr}
\end{figure}

We remark that the forces derived from a three-body potential have a
geometrical interpretation analogous to those for a two-body
potential.  In the two-body case, the force on each particle is
directed along the line that contains the two particles.  In the
three-body case, the force on each of the three distinct particles 
is directed along a radial line through the particle and the center 
of the circle that contains all three of the particles 
(cf.\ Figure \ref{FigThr}).  

\subsection{Other distances for curves}

The three-point circumradius function leads to a notion of distance
between pairs of points that contains geometrical information in
addition to the standard distance.  For example, if $\bx=\bq(s)$,
$\by=\bq(\sigma)$ and $\bz=\bq(\tau)$ are three distinct points on
$\esC$, then one can associate a
distance to $\bx$ and $\by$ according to
$$
\Pt(\bx,\by)= \lim_{\tau\to\sigma}r(\bx,\by,\bz)
= \frac{|\bx-\by|}{2|\sin\theta_{\bx\by'}|}
$$
where $\theta_{\bx\by'}$ is the angle between the vector
$\bx-\by\ne\bzero$ and the tangent vector to $\esC$ at $\by$. The
function $\Pt(\bx,\by)$ can be identified as the radius of the unique
circle through $\bx$ that is tangent to $\esC$ at $\by$.  Notice that
the function $\Pt(\bx,\by)$ will typically not be symmetric since the
circle through $\by$ and tangent at $\bx$ need not have the same
radius as the circle through $\bx$ and tangent at $\by$.  We will
refer to $\Pt(\bx,\by)$ as the point-tangent function. As a matter of
convention, we consider it to be a particular three-point function,
because it corresponds to a limit of the three-point circumradius
function.

The point-tangent function $\Pt(\bx,\by)$ shares two 
important properties with the circumradius function 
$r(\bx,\by,\bz)$.  First, the coalescent limit 
$\sigma\to s$ is just the local radius of curvature, 
namely
$$
\lim_{\sigma\to s}\Pt(\bx,\by) = \rho(\bx).
$$
Second, whenever different neighborhoods of a curve come together to 
form a self-intersection, there are points $\bx$ and $\by$ for which
$\Pt(\bx,\by)$ and $\Pt(\by,\bx)$ are both equal to half of the distance
of closest approach.  As a consequence of these two properties, a
general class of self-interaction energies involving the point-tangent
function as argument can be defined as
$$
E[\esC] = \iint V(\Pt(s,\sigma))\;\dC_s\;\dC_\sigma
$$
where $\Pt(s,\sigma) = \Pt(\bq(s),\bq(\sigma))$ and $V$ 
is a general potential function as before.  The energy 
$E[\esC]$ is well-defined for any simple curve $\esC$. 

Other distances for curves based on four points, or limits thereof,
can also be considered.  For example, to any four distinct points
$\bw$, $\bx$, $\by$ and $\bz$ on a simple curve $\esC$ one can
associate the four-point distance $R(\bw,\bx,\by,\bz)$ defined as the
radius of the smallest sphere that contains all four points. Usually
there will be such a unique sphere, but if all four points happen to
be co-circular, there are many spheres passing through them with the
smallest having the radius of the circle through the four points. Any
limit of the spherical radius function as the four points coalesce
into one is always positive because it is greater than or equal to the
radius of curvature of the curve at the limit point.  If the
coalescence point has non-zero torsion, one obtains the radius of 
the {\em osculating sphere} in the limit, see for example 
\cite[p.\ 25]{STRUIK}.  Moreover, just as for the three-point functions 
$r$ and $\Pt$, whenever different parts of a curve come together to form 
a self-intersection, there are points for which $R$ is equal to half
of the distance of closest approach. (But now the possibility
of end-point effects must be explicitly excluded, as for example if the
curve is closed.)  Similar results hold for the symmetric
tangent-tangent distance function $\Tt(\bx,\by)$ defined to be the
radius of the smallest sphere containing $\bx$ and $\by$, that is also
tangent to $\esC$ at both these points.

\section{Self-interactions of surfaces}
\label{SecSurf}

Self-avoiding surfaces have been a subject of much interest and study
in diverse disciplines ranging from mathematics to biology (see for
example \cite{NELSON}, \cite{WIESE}, \cite{WIESEb}--\cite{KROLL}).
Our objective is therefore to extend to the case of surfaces, our
construction of multi-point distances that can lead to non-singular
repulsive energies that preclude self-intersection.  Our first
conclusion is that for surfaces the simplest two-point (Euclidean),
three-point (circular), and four-point (spherical) distances all share
a similar problem: they cannot distinguish between proximity due to
continuity and proximity due to geometry. We will demonstrate,
however, that the notion of point-tangent distance for a curve has a
natural counterpart for surfaces that can make this distinction, and
which provides a suitable argument $\Varg$ for a self-interaction
potential.  Throughout our developments a surface $\esS$ will mean a
(twice continuously differentiable) function $\bp(u)\in\Reals^3$ of a
variable $u\in A\subset\Reals^2$, and a surface will be called simple
if it has no self-intersections; that is, $\bp(u)=\bp(v)$ only when
$u=v$.

\subsection{$n$-point distances for surfaces ($n=2,3,4$)}

Just as for the case of curves, an intuitive approach to the
construction of a self-interaction energy for a simple surface $\esS$
is to consider a repulsive potential dependent upon the pairwise or
two-point distance function $\Edist(u,v)=|\bp(u)-\bp(v)|$ as argument.
As before, while $\Edist(u,v)$ has the desirable feature that it tends to 
zero as different neighborhoods of $\esS$ approach a self-intersection, 
it also tends to zero as $u\to v$ by continuity, and thus leads to 
singular interaction potentials.

In contrast to the case of curves, one may show that for surfaces both
the three-point function $r(u,v,w)$ and the four-point function
$R(t,u,v,w)$ introduced earlier, also lead to singular
self-interaction potentials.  In particular, these functions tend to
zero by continuity in the coalescent limit.  This conclusion may be
established as follows.  Consider any fixed point $\bx$ of $\esS$ and
let $\Sigma_\varepsilon(\bx)$ be the sphere of radius $\varepsilon$
centered at $\bx$.  For each sufficiently small $\varepsilon>0$ the
intersection $\Sigma_\varepsilon(\bx)\cap\esS$ of the sphere and
surface is a curve. The radius $r$ of the circle through any three
points on this curve satisfies $r\le\varepsilon$ (because the circle
lies on the sphere of radius $\varepsilon$), and similarly the radius
$R$ of the smallest sphere through any four points on this curve
satisfies $R\le\varepsilon$.  Thus, by considering the limit
$\varepsilon\to 0$, we can find sequences of three distinct points in
the surface for which $r\to 0$, and sequences of four distinct points
for which $R\to 0$.

\subsection{Tangent-point distance for surfaces}

For surfaces, the argument of a non-singular self-interaction potential 
can be obtained by passing directly to tangent-point distances.  
In particular, to any two distinct points $\bx$ and $\by$ 
of a simple surface $\esS$ we may associate the distance
$$
\Tp(\bx,\by) = \frac{|\by-\bx|^2}{2|\bn_\bx\cdot(\by-\bx)|}
$$
where the vector $\bn_\bx$ is either of the two unit normals to the
surface $\esS$ at the point $\bx$.  The function $\Tp(\bx,\by)$ can be
identified as the radius of the unique sphere through $\by$ and
tangent to $\esS$ at $\bx$.  When $\by$ happens to be in the tangent
plane to $\esS$ at $\bx$, the sphere itself degenerates into a plane
and $\Tp(\bx,\by)$ becomes infinite.  Notice that $\Tp(\bx,\by)$ need
not be symmetric since the sphere through $\bx$ and tangent at $\by$
need not have the same radius.  We refer to $\Tp(\bx,\by)$ as the
tangent-point function for surfaces. Moreover, we consider it to be a
particular four-point function since the tangent plane to $\esS$ at
$\bx$ may be constructed through a coalescent limit of three points.

The tangent-point function $\Tp(\bx,\by)$ enjoys various 
properties analogous to those of its counterpart 
defined for curves. For example, consider any curve $\bq(s)$
in $\esS$  such that $\bq(0)=\bx$ and let $\bt_\bx=\bq'(0)$.
Then
$$
\lim_{s\to 0}\Tp(\bx,\bq(s)) 
= \frac{1}{|\bn_\bx\cdot\bq''(0)|}
= \rho(\bx,\bt_\bx)
$$
where $\rho(\bx,\bt_\bx)$ is the absolute value of the local normal
radius of curvature to $\esS$ at $\bx$ in the direction $\bt_\bx$.
Thus, in coalescent limits $\by\to\bx$, $\Tp(\bx,\by)$ may assume
limiting values between the maximum and minimum of $\rho(\bx,\bt_\bx)$
over directions $\bt_\bx$ with $\bx$ fixed.  When different parts of
$\esS$ come together to form a self-intersection, there are points
$\bx$ and $\by$ for which $\Tp(\bx,\by)$ and $\Tp(\by,\bx)$ are both
equal to half of the distance of closest approach of the surface to
itself.

While $\Tp(\bx,\by)$ need not be continuous when $\by=\bx$, one may 
unambiguously consider self-interaction energies for surfaces
of the form
$$
E[\esS] = \iint V(\Tp(u,v))\;\dS_u\;\dS_v
$$
where $\Tp(u,v) = \Tp(\bp(u),\bp(v))$ and $V$ is a general potential 
function.  As in the case of a spaghetto, the choice of a potential 
of the form introduced in Figure \ref{FigOne} could be used to model 
a surface of finite thickness.  Similar ideas could be used for the 
description of triangulated or discretized random surfaces.  In 
particular, one may discretize the energy introduced above to obtain 
a self-interaction energy for the triangulated surface.  Here three 
vertices of each triangle could be used to define a tangent plane, 
which could then be used to evaluate the tangent-point function $\Tp$.

We remark that  M{\"o}bius-invariant energies for self-avoiding surfaces 
involving pairs of spheres, each tangent to the surface, are discussed in
\cite{KUSNER}.  However, energies based on the radius of a single such 
sphere are not discussed.

\section{Summary}
\label{SecSumm}

Curves and surfaces with an associated finite thickness can be used 
as continuum models of strands (such as a spaghetto) and sheets (such 
as this page).  We find that non-singular self-interaction energies 
for such self-avoiding continuous systems cannot depend on the usual 
pairwise distance.  However, by using certain many-body potentials, 
we are able to construct self-interaction energies which are non-singular 
and which include a mesoscopic length scale for the physical thickness.
Each  many-body potential is a function of the radius of a suitably 
chosen circle or sphere.  Our energies provide simple continuum 
models of strand and sheet systems that can be used to study the 
equilibrium and non-equilibrium statistical mechanics of distinct 
phases at different temperatures, along with phase transitions.

We conclude by highlighting an application of our work in the context
of the protein problem.
Small globular proteins are linear chains of amino acids which, under
physiological conditions, fold rapidly and reproducibly in a
cooperative manner into their native state conformations \cite{Anfinsen}.
These conformations are somewhat compact structures corresponding to 
the minima of an effective energy.
Furthermore, for proteins, form determines function and, yet, the
total number of distinct native state folds is believed to be only of
the order of a few thousand \cite{Chothia} and are made up of
helices, hairpins or sheets.  
An important issue in the protein field is to elucidate
the bare essentials that
determine the novel phase adopted by biopolymers such as proteins.
Unfortunately, polymer science, which is a mature and technologically
important field, does not provide an answer to this question.  The 
standard, simple model of a polymer chain is one of  tethered hard
spheres -- in the continuum limit, self-avoidance  in this model
is captured by a singular pairwise potential as in the second term 
on the right hand side of equation (\ref{Ham_2}).  Such a model 
with an additional effective attraction arising from the
hydrophobicity or mutual aversion of certain amino acid residues to the 
solvent (water), fails to account for the protein native structure phase
on several counts.  First, a generic compact polymer phase has
many conformations which neither provide for specificity nor are
kinetically readily accessible.  Proteins have a limited number of
folds to choose from for their native state structure and the energy
landscape is vastly simpler.  Second, the structures in the polymer
phase are not especially sensitive to perturbations and are thus not
flexible and versatile as protein native state structures are in order
to accommodate the dizzying array of functions that proteins perform.

Recent work \cite{Banavar}, built on the ideas presented here, has
pointed out a crucial missing ingredient in the simple model of a
polymer chain of tethered hard spheres.  Strikingly, the simple
physical idea of a chain viewed, instead, as a tube leads to several
dramatic consequences.  We are conditioned to think of objects as
spheres and the effective interactions between them as being pair-wise
in nature.  This bias in our thinking arises from our everyday
experience with unconstrained objects.  However, when one deals with
objects tethered together in a chain, some of the old notions need to
be discarded.  For example, a pairwise interaction only provides
information regarding the mutual distance between two interacting
particles but does not have any contextual information regarding how
far apart the two particles are along the chain.  Our work here shows
that in order to capture the constraints imposed by a tube geometry
associated with a discrete chain, the conventional notion of pairwise
interactions between particles has to be augmented by appropriately
chosen three-body interactions to capture the steric constraints
imposed by the tube.

Such a tube-like description of a chain leads to many of the standard
results of polymer physics when the tube thickness is small compared
to other length scales in the problem \cite{Banavar}.  However, when
the tube size becomes comparable to the range of the effective
attractive interactions resulting from the hydrophobicity, the novel
phase populated by biopolymers results \cite{Banavar}.  This
self-tuning of length scales occurs naturally for proteins because the
steric effects \cite{Rama,Linus} associated with the backbone and side
chain atoms of the amino acids, on the one hand, lead to a tube like
description and control the tube thickness and the same side chains,
on the other hand, have an effective attractive interaction which is
at an Angstrom scale due to the screening effects of the water.

It has been demonstrated \cite{Banavar} that this novel phase has
several characteristics of the phase populated by protein native
structures including the ability to expel water from the hydrophobic
core, a vast simplification in the energy landscape with relatively
few putative native state structures, a prediction that helices,
zig-zag hairpins and sheets of the correct geometry are the building
blocks of protein structures, a simple explanation of the cooperative
nature of the folding transition of globular proteins and an
explanation of why protein structures are flexible and versatile.
Thus the idea of a tube and the use of appropriate many-body
potentials introduced here not only lead to a better description of a
polymer but allows one to bridge the gap between polymer science and
protein science and provides, for the first time, a framework based on
geometry for an understanding of the common character of all globular
proteins \cite{Banavar}.  We look forward to similar applications of
our new ideas on sheets of non-zero thickness.

%--------------
% References
%--------------

%-----------------
% Acknowledgements
%-----------------

\noindent
{\bf Acknowledgments.}  
We are indebted to Philip Anderson and Leo Kadanoff for valuable 
comments on the manuscript.  The following generous support is
gratefully acknowledged: JRB the Penn State MRSEC under NSF 
grant DMR-0080019 and NASA, OG the US National Science Foundation 
under grant DMS-0102476, JHM the Swiss National Science Foundation, 
and AM the MURST Cofin'99. 

\vfill\eject

%-----------------
% Figures
%-----------------

\vfill\eject

%%%%%%%%%%%%%%%%%%%%%%%%%%%%%%%%%%%%%%%%%%%%%%%%%%%%%%%%%%%%%%%%%%%%%%%
% End of the article
%%%%%%%%%%%%%%%%%%%%%%%%%%%%%%%%%%%%%%%%%%%%%%%%%%%%%%%%%%%%%%%%%%%%%%%
\end{document}